# Consensus Recommendations for Hyperpolarized [1-$^{13}$C]pyruvate MRI Multi-center Human Studies

Shonit Punwani 1,2, Peder EZ Larson 3, Christoffer Laustsen 4, Jan VanderMeulen 5, Jan Henrik Ardenkjær-Larsen 14, Adam W. Autry 3, James A. Bankson 6, Jenna Bernard 3, Robert Bok 3, Lotte Bonde Bertelsen 4, Jenny Che 3, Albert P. Chen 7, Rafat Chowdhury 1,2, Arnaud Comment 8, Charles H. Cunningham 9,10, Duy Dang 3, Ferdia A Gallagher 11, Adam Gaunt 12, Yangcan Gong 1,2, Jeremy W. Gordon 3, Ashley Grimmer 11, James Grist 13,28, Esben Søvsø Szocska Hansen 4, Mathilde Hauge Lerche 14, Richard L. Hesketh 15, Jan-Bernd Hoevener 16, Ching-Yi Hsieh 17, Kayvan R. Keshari 18, Sebastian Kozerke 19, Titus Lanz 20, Dirk Mayer 21,22, Mary McLean 11,23, Jae Mo Park 24,25,26, Jim Slater 3, Damian Tyler 13,27,28, Jean-Luc Vanderheyden 29, 30,Cornelius von Morze 31, Fulvio Zaccagna 11,13,32,33,34, Vlad Zaha 24,35, Duan Xu 3, Daniel Vigneron 3 and the HP 13C MRI Consensus Group

Affiliation:
1 Department of Radiology, University College London Hospital NHS Foundation Trust, London, UK.
2 Centre for Medical Imaging, University College London, London, UK.
3 Department of Radiology and Biomedical Imaging, University of California, San Francisco, CA 94143, USA.
4 The MR Research Center, Department of Clinical Medicine, Aarhus University, Aarhus,Denmark.
5 Department of Health Services Research & Policy, London School of Hygiene & Tropical Medicine, London, UK.
6 Department of Imaging Physics, The University of Texas MD Anderson Cancer Center, Houston, TX, USA.
7 GE Healthcare, Toronto, Ontario, Canada.
8 GE HealthCare, Pollards Wood, Nightingales Lane, Chalfont St Giles, Buckinghamshire HP8 4SP, United Kingdom
9 Physical Sciences, Sunnybrook Research Institute, Toronto, Ontario, Canada.
10 Department of Medical Biophysics, University of Toronto, Toronto, Ontario, Canada.
11 Department of Radiology, University of Cambridge and Cambridge University Hospitals NHS Foundation Trust, Cambridge, UK.
12 GE Healthcare, San Francisco, California, USA.
13 Oxford Centre for Clinical Magnetic Resonance Research, University of Oxford, Oxford, UK.
14 Department of Health Technology, Technical University of Denmark,2800, Kgs. Lyngby,Denmark.
15 Cancer Research UK Cambridge Institute, University of Cambridge, Cambridge, United Kingdom.
16 Department for Radiology and Neuroradiology, Molecular Imaging North Competence Center (MOIN CC), Section Biomedical Imaging, University Hospital Schleswig-Holstein - Campus Kiel, Kiel University, Germany.

17 Research Center for Radiation Medicine, Chang Gung University, Taoyuan, Taiwan, and Department of Medical Imaging and Intervention, Chang Gung Memorial Hospital at Linkou, Taoyuan, Taiwan.
18 Department of Radiology, Memorial Sloan Kettering Cancer Center, New York, NY, USA; Molecular Pharmacology Program, Memorial Sloan Kettering Cancer Center, New York, NY,USA.
19 Institute for Biomedical Engineering, University and ETH Zurich, Zurich, Switzerland.
20 RAPID Biomedical GmbH, 97222 Rimpar, Germany.
21 Department of Diagnostic Radiology and Nuclear Medicine, University of Maryland School of Medicine, Baltimore, Maryland, USA.
22 University of Maryland Marlene and Stewart Comprehensive Cancer Center, Baltimore,Maryland, USA
23 Cancer Research UK Cambridge Institute, University of Cambridge, Cambridge, UK.
24 Advanced Imaging Research Center, The University of Texas Southwestern Medical Center, Dallas, Texas, USA.
25 Radiology, The University of Texas Southwestern Medical Center, Dallas, Texas, USA.
26 Electrical and Computer Engineering, The University of Texas at Dallas, Richardson, Texas,USA.
27 Department of Physiology, Anatomy, and Genetics, University of Oxford, Oxford, UK.
28 Department of Radiology, Oxford University Hospitals, Oxford, UK.
29 GE Healthcare, Menlo Park, California, USA.
30 JLVMI Consulting LLC, Dousman, WI, USA
31 Department of Radiology, Washington University, St. Louis, Missouri, USA.
32 Department of Medicine, University of Cambridge, United Kingdom.
33 Department of Imaging, Cambridge University Hospitals NHS Foundation Trust, Cambridge Biomedical Campus, Cambridge, United Kingdom.
34 Investigative Medicine Division, Radcliffe Department of Medicine, University of Oxford,Oxford, United Kingdom.
35 Internal Medicine, The University of Texas Southwestern Medical Center, Dallas, Texas, USA.

This work was supported and endorsed by the ISMRM Hyperpolarized MR Study Group

Word count: 6606 (excluding title, abstract reference, table and captions)

# Abstract

Magnetic resonance imaging of hyperpolarized (HP) [1-$^{13}$C]pyruvate allows in-vivo assessment of metabolism and has translated into human studies across diseases at 15 centers worldwide. Consensus on best practice for multi-center studies is required to develop clinical applications. This paper presents the results of a 2-round formal consensus building exercise carried out by experts with HP [1-$^{13}$C]pyruvate human study experience. Twenty-nine participants from 13 sites brought together expertise in pharmacy methods, MR physics, translational imaging, and data-analysis; with the goal of providing recommendations and best practice statements on conduct of multi-center human studies of HP [1-$^{13}$C]pyruvate MRI.

Overall, the group reached consensus on approximately two-thirds of 246 statements in the questionnaire, covering ‘HP $^{13}$C-Pyruvate Preparation’, ‘MRI System Setup, Calibration, and Phantoms’, ‘Acquisition and Reconstruction’, and ‘Data Analysis and Quantification’.

Consensus was present across categories, examples include that: (i) different HP pyruvate preparation methods could be used in human studies, but that the same release criteria have to be followed; (ii) site qualification and quality assurance must be performed with phantoms and that the same field strength must be used, but that the rest of the system setup and calibration methods could be determined by individual sites;(iii) the same pulse sequence and reconstruction methods were preferable, but the exact choice should be governed by the anatomical target; (iv) normalized metabolite area-under-curve (AUC) values and metabolite AUC were the preferred metabolism metrics.

The work confirmed areas of consensus for multi-center study conduct and identified where further research is required to ascertain best practice.

# Introduction

Magnetic resonance imaging of hyperpolarized (HP) [1-$^{13}$C]pyruvate allows for the real time assessment of in vivo metabolism. In 2013, the first human study with HP [1-$^{13}$C]pyruvate was reported in patients with prostate cancer (1). Since then, there have been over 100 reports of human applications including multiple types of cancer (e.g. prostate, brain, breast, kidney, pancreas), liver disease, ischemic heart disease, neurodegenerative disease, diabetes and cardiomyopathies (2). Whilst demonstrating proof of concept, feasibility, and insights into biology, none of these potential applications has been developed further into an approved clinically-reimbursable application.

Development of new clinical imaging methods requires a series of steps, via sequential and progressive studies, such as outlined for cancer in the Cancer Research UK (CRUK) Biomarker Roadmap for Cancer (3,4). The promising results with HP [1-$^{13}$C]pyruvate to date place the technology at the cusp of clinical translation. Fifteen centers around the world have so far conducted human studies, with a total of over 1200 participants having been scanned. Moreover, here is also active development of hyperpolarization technologies, which has the potential to improve support of large, multi-center human studies and clinical trials.

Standardizing methodology of HP [1-$^{13}$C]pyruvate MRI is now necessary in preparation for such studies. Therefore we conducted a formal consensus exercise focused on establishing best practice (and identifying areas for further research) needed to assist researchers in the planning of multi-center human studies and clinical trials, where generation of datasets across centers will be required.

Selected consensus group members participated in 2 rounds of voting, the first one completed remotely individually and the second one after discussion of the results of the first round during a face-to-face meeting. The group was given the following goal and scope of this process:

**Goal**: The results from this formal consensus exercise will be used to provide recommendations and best practice statements for how to execute human studies with HP [1-$^{13}$C]pyruvate and to identify areas where additional research should be conducted.

**Scope**: Please state your agreement with the consensus statements using your expertise to 'recommend' to researchers who are currently performing human studies with dDNP [1-$^{13}$C]pyruvate or planning to perform these studies.

This manuscript presents the consensus results in terms of agreement with consensus statements and describes discussions that occurred during the face-to-face meeting. The specific methodologies that the consensus statements refer to are not comprehensively described or referenced in this paper, as this information can be found in the prior "Position Paper" on current methods for HP [1-$^{13}$C]pyruvate MRI human studies that was also authored by the HP $^{13}$C MRI Consensus Group (2).

# Materials and Methods

## The consensus method

The RAND-UCLA Appropriateness Method (RAM) was chosen as the most appropriate for our objectives and has been extensively used for development and utilization of new imaging technologies (5,6). RAM includes a process of mailed and face-to-face consensus rounds. Its remit is to enable recommendations in topic areas where there is little or poor-quality evidence. A scale for scoring is used to indicate the level of agreement or disagreement with each statement under discussion on a 9-point scale. The RAM User's Manual was followed throughout the process (5).

## Group selection

Leads [JS, APC, JLV, DM, MM, SP, CC, CV, JWG, CL, JAB, DX, RB, JH] for the various sections to be discussed within the consensus were established during the prior literature review (2). The leads recommended participants for inclusion within the consensus to the core consensus group (SP, PL, CL, DV, DX). Members were eligible to be included if they met the following criteria:

(i) had previously participated in a human [1-$^{13}$C]pyruvate study,
(ii) held either a Faculty, Physician, Pharmacist, Staff Scientist or Post-Doctoral position and

(iii) were available to participate in an in-person consensus meeting held in San Francisco on April 3rd 2024, just prior to the 2024 Hyperpolarized C-13 Technology Workshop.

The consensus group included several (APC, AC, AG, TL and JLV) experienced industry scientists and engineers specializing in HP [1-$^{13}$C]pyruvate studies, many of whom migrated from academia to industry while continuing to work in the field. They contributed valuable insights to HP [1-$^{13}$C]pyruvate human studies. They have no commercial roles and were prohibited from voting on vendor-specific questions.

The chair, Jan van der Meulen, is a clinical epidemiologist with experience using formal consensus methods to develop clinical guidelines. Invited participants covered a diverse set of roles in the HP $^{13}$C MRI field, including pharmaceutical preparation methods, MR physics, translational imaging, and data-analysis. A total of 32 participants completed the round one electronic survey. A final count of 29 participants participated in the face-to-face meeting, representing a total of 13 HP $^{13}$C MRI sites engaged in human work. These group members consisted of 6 clinicians, 4 pharmacists and 22 MR Scientists. Details for members and their relevant experiences are given in Supplementary Material Table 1.

## Construction of the questionnaire

A questionnaire containing 246 statements was developed between August 2023 and December 2023. The initial draft was compiled by the leads. Statements were then further refined by the consensus leads with the core consensus group. Statements were designed to address the mandatory and preferred requirements for [1-$^{13}$C]pyruvate HP multi-center human studies. The statements covered four topics in HP C-13 MRI: Hyperpolarized C-13 Pyruvate Preparation; MRI System Setup, Calibration & Phantoms; Acquisition & Reconstruction; and Data Analysis and Quantification (Figure 1). These topics were identified in the systematic literature review (2).

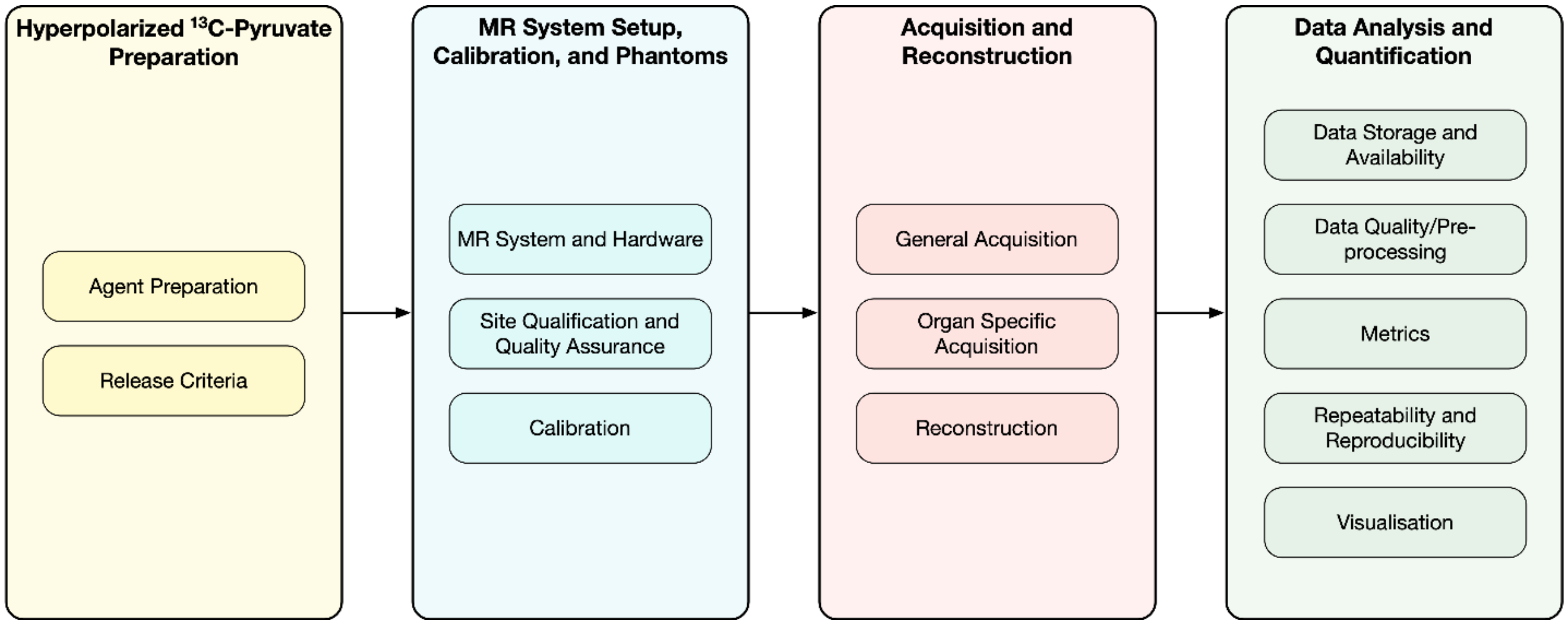

*Figure 1: Illustration of the HP [1-$^{13}$C]pyruvate MRI human study process, including the 4 major areas covered in this paper: Hyperpolarized [1-$^{13}$C]pyruvate preparation, MRI system setup and calibration, Acquisition and Reconstruction, and Data Analysis and Quantification.*

There were three styles of statements that participants were asked to score:

The first style of statement was applied to more established practices that the leads felt that there was sufficient evidence of performance to potentially make strong ‘have to’ recommendations.  For these statements, we asked whether sites ‘have to’ or it is ‘preferred’ they follow a certain practice in multi-center studies. Participants were asked to score their agreement with the same statement twice, initially using the wording ‘have to' and then 'preferred'.

The second style of statements was applied to practices where the leads felt that evidence was likely too limited to make a ‘have to’ recommendation. In this case, participants were asked to score their agreement with the same statement twice, initially using the wording 'viable' and then 'preferred'.

The final style of statements directly asked the consensus group to rate their agreement with statements about 'best practice'. These best practice statements were used to ask about more specific practices for HP $^{13}$C studies, and included practices that are likely to vary across different multi-center studies.

There is a complete list of statements in Supplementary Table 2, and the different styles are evident by the specific word choices in the statements as described above.

## First-round questionnaire completion before meeting

Participants were asked to have read "Current methods for hyperpolarized [1-$^{13}$C]pyruvate MRI human studies", which is a systematic literature review of current methodology and practices utilized in HP $^{13}$C human clinical studies (2).

All agreement with the statements was scored on a scale of 0-9 and were processed and grouped as follows: (1,2,3 = disagree), (4,5,6 = uncertain), (7,8,9 = agree). Voters were allowed to skip statements if they felt they were not within their area of expertise (0 = I don't know enough about the topic to answer). The questionnaire was distributed electronically, and in the first round each participant was given 2 weeks to submit their agreement scores. The same scoring scale was used during the subsequent face-to-face meeting.

## Face-to-face meeting format

The face-to-face meeting was conducted in-person on April 3, 2024, prior to the start of the 2024 HP C-13 Technology Workshop at UCSF in San Francisco, USA, which was held on April 4-5, 2024.

During the face-to-face meeting, all participants were provided with a list of all statements and their original agreement scores from the first round. The summative first-round scores were then reviewed statement by statement, by displaying a bar chart of the score distribution, total number of members scoring the question, and whether or not scores reached a consensus. The chair (JVM) then hosted a timed discussion that allowed as many participants as possible to describe their arguments for and against agreement with the statement. Following discussion of individual statements, participants were asked to re-score the statement.

## Interpretation of results

The results were interpreted according to the RAM User's Manual. The outcome category (1-3 disagree, 4-6 uncertain or 7-9 agree) was defined by the median score across all non-zero responders. 'Consensus' was defined to be having 66% or more of the responses with non-zero scores in any one category among (disagree, uncertain and agree). Only those statements scored non-zero by at least 14 group members (approximately 50% of the panelists within our group) were considered for consensus recommendations.

# Results

Please see Supplementary Table 2 for a list of all statements and a summary of the results from the second round.  At the end of the process, there was consensus for 146/235 (62.1%) of the statements scored; increased from 133/246 (54.1%) at the end of the first round (Supplementary Table 3).  During the face-to-face meeting, 18 statements were found to be either inadequate or ambiguous and the statement was rephrased to improve clarity, while an additional 11 statements were removed, with all changes being made after full agreement by group members (Supplementary Table 4).

## Hyperpolarized $^{13}$C-Pyruvate Preparation

In this section, we examined how HP [1-$^{13}$C]pyruvate doses are prepared.  This process requires that [1-$^{13}$C]pyruvate becomes hyperpolarized and then retains that hyperpolarization in the liquid-state to become an agent that can safely and effectively be injected into human subjects.  The statements covered two major topic areas of agent preparation, and the release criteria that must be met prior to injection.  Note that all current methods for human dose preparation use dissolution dynamic nuclear polarization (dDNP), thus many of these recommendations are specific to that approach. Key consensus recommendations are summarized in Table 1.

### Agent Preparation

Regarding how HP [1-$^{13}$C]pyruvate is made in a multi-center study, there was consensus reached that sites do not have to follow the same production style (17/25 , 68%), and following from this that they do not have to use the same standard operating procedures (SOPs) (20/26, 77%) or the same environment and facilities (19/26, 73%). Currently, there are two major production styles used in approved clinical studies: Terminal Sterilization, which relies on an endline sterilizing filter, like PET radionuclide dose preparation and which is more common in the US, and Aseptic Preparation, which more closely follows sterile pharmaceutical compounding guidelines and is more common elsewhere in the world for HP $^{13}C$ dose preparation (2). Aseptic technique follows the United States Pharmacopeia (USP) 797 regulations for pharmacy sterile compounding. Terminal sterilization follows the manufacturing process outlined in USP 823 (Radiopharmaceuticals for Positron Emission Tomography [PET]—Compounding) and Title 21 Code of Federal Regulations (CFR) Part 212 (Current Good Manufacturing Practice for Positron Emission Tomography Drugs). The primary distinction between these approaches lies in the sterility requirements for ingredients and materials. Aseptic sterile compounding mandates that all ingredients and materials be sterile from the outset, and the use of a sterility filter is not required. In contrast, terminal sterilization does not necessitate sterility of ingredients and materials before filtration; instead, the drug is sterilized using a sterility filter. Everything post filtration needs to be sterile which is the drug closure. In addition, a filter integrity test is performed before the drug is administered. This result reflects the discussion in the meeting that participants did not find how HP [1-$^{13}C$]pyruvate is made to be relevant in the context of a multi-center study, provided that a standard dose specification can be achieved and it can be validated to be safe.

## Release Criteria

Regarding the release of the HP agent for injection, there was consensus reached that all sites in a multi-center study have to use the same release criteria (21/26, 81%). This was widely agreed upon, and it was noted that this follows how PET is used in multi-center studies where, for example, a minimum activity is required for release. The discussed goals for having release criteria were 1) safety as the most important, 2) providing metrics of the fitness of equipment and processes, and 3) being practical, especially given the rapid decay of the HP [1-$^{13}C$]pyruvate. Similar to the production discussion, there was consensus reached that sites do not have to use the same quality control (QC) equipment for the release criteria (17/24, 71%), but with the assumption that each site will need to validate their own processes.

Of the specific release criteria, there was unanimous support that the release criteria has to include measurements of pH and the residual electron paramagnetic agent (EPA) concentration (26/26,100%). There was also consensus that the release criteria has to include pyruvate concentration (25/26, 96%), temperature (23/26, 88%), and volume of the HP [1-$^{13}C$]pyruvate dose (21/26, 81%). There was no consensus that polarization measurements have to be included (14/25, 56%). There was discussion highlighting that polarization measurements would be desirable, but group members indicated that there was a lack of confidence in current measurement methods and that it is difficult to measure quickly. The lack of consensus on polarization measurement necessitates further discussion in follow-up meetings, which would consider advancements in hardware and polarization methodologies. Given current technological capabilities, solid-state polarization build-up may serve as a surrogate for liquid-state polarization. Additionally, data quality, including the impact of polarization, should be evaluated using image signal-to-noise ratio (SNR). The discussion also noted that there are strong correlations between various release criteria (e.g., pH and pyruvic acid concentration are strongly linked), but this redundancy was also considered desirable as it increases safety and confidence.

## Summary

Consensus was reached by the group to recommend a minimum set of parameters related to the HP agent that have to be recorded and reported by HP [1-$^{13}$C]pyruvate multi-center human studies, which are shown in Table 5.

In summary, the group agreed that they are not concerned about exactly how HP [1-$^{13}$C]pyruvate doses are made, so different processes can be used, which is advantageous given that different sites currently use and have regulatory approvals for the differing production styles of Terminal Sterilization and Aseptic Preparation. The group was in agreement in requiring the same release criteria to serve as a way to both ensure safety as well as ensure similar performance across sites regardless of the production style. As we move towards multi-center studies, the specific release criteria would need to be defined. How we work with regulatory environments internationally is an open question, but we encourage the community to try to put forth a united front on best practices and look for opportunities to lead the decision-making regarding a safe yet practical set of regulatory requirements.

*Table 1: Summary of Hyperpolarized $^{13}$C-Pyruvate Preparation recommendations. These are based on the currently available dDNP methods for human dose preparation.*

| **Summary of Hyperpolarized $^{13}$C-Pyruvate Preparation recommendations** |
|---|
| Sites do not have to follow same production style, QC, or other processes |

Sites have to use the same release criteria to ensure safety and performance, and this should include pH, EPA concentration, pyruvate concentration, temperature and volume

## MRI System Setup, Calibration, and Phantoms

In this section, we examined the characteristics and calibration of the MRI system and hardware, site qualification and quality assurance (QA), and phantoms. HP [1-$^{13}$C]pyruvate human studies require $^{13}$C capability in the RF system and coils. Specialized pre-scan calibration methods are also required, as most MRI methods rely on $^{1}$H signals in the body for these calibrations but there are only very small signals available from natural abundance $^{13}$C. Custom phantoms have also been used to test the $^{13}$C capability of the system, test the performance of HP [1-$^{13}$C]pyruvate acquisition and reconstruction methods, and perform calibration measurements. Key consensus recommendations are summarized in Table 2.

### MRI System and Hardware

Regarding the MRI system, in most regards the group did not want to require a certain set of characteristics such as gradient system capabilities, RF system capabilities, or scanner vendor. The only exception was that there was consensus on requiring the same $B_0$ for inclusion in a multi-center study (19/24, 79%). This was not seen as a limitation, as all published human studies have used 3T MRI systems. This may be driven by the fact that major MR vendors only offer non-$^{1}$H imaging capability on field strength of 3T or above, and there is limited evidence regarding the performance of human studies at other field strengths (7).

Regarding RF coils, there was consensus preferring that the $^{13}$C transmit and receive RF coils at institutions participating in a multi-center study have the same fundamental geometry design (22/23, 96%). There was also consensus preferring that the same manufacturer and model be used (20/24, 83%). However, these two questions asked with the ‘have to’ (ie. mandatory) prefix did not reach consensus. There was consensus that some $^{1}$H images have to be acquired with the subject in the same position used for the $^{13}$C images (16/24, 67%). The purpose of this is to provide an anatomical reference for the HP data, and this recommendation allows for acquiring additional $^{1}H$ images with repositioning, for example to use higher performance $^{1}H$ coils.

## Site Qualification and Quality Assurance

For both site qualification and quality assurance (QA), there was consensus that these have to be done.  Using static phantoms with thermally polarized solutions, as opposed to using hyperpolarized solutions, reached consensus as the preferred option (25/25, 100%). There was discussion regarding the use of dynamic, hyperpolarized phantoms, but there was no consensus on whether they are or are not a viable option for site qualification. There was considerable debate about using traveling humans as phantoms, with agreement scores showing no consensus as a viable approach and the discussion raising issues of biological variability and logistical challenges.

## Calibration

Finally, calibration of the RF power, center frequency, receiver gains, and shimming were discussed. For the RF power, center frequency, and shimming, there was consensus to prefer the same methods but that this should not be required to be the same across the sites. For RF power, both automatic optimization and using manual optimization (e.g. manually performed power sweeps) were agreed to be viable methods.  Automatic methods received the strongest support as a preferred method (21/21, 100%). Using predetermined values from phantom measurements for calibration is common practice, but there was no consensus that this is a viable method. For center frequency, using the 1H water frequency to estimate the 13C frequency based on the 1H and 13C gyromagnetic ratios and relative chemical shifts emerged as the clearly preferred method (23/24, 96%). For receiver gains, there was no consensus to prefer or require the same methods be used across sites. Using previous *in vivo* measurements for receiver gains was the only consensus viable option (19/22, 86%). For shimming, conventional auto-shimming methods on proton signals reached consensus to be viable and were the most preferred method. Manual shimming methods did not reach consensus as viable methods. No consensus was reached on requiring or preferring higher order shimming.

## Summary

In summary, it was agreed that multi-center studies should use the same $B_0$, but that differences in vendor, scanner specifications, and RF coils would be acceptable. The discussion showed that there are unmet needs in phantoms for HP $^{13}$C studies. Calibration is challenging due to the relatively small natural abundance $^{13}$C signal *in vivo*, and the consensus results indicate that RF power and receive gain calibration remains challenging. The results for shimming and center frequency calibration, however, showed that using $^{1}$H signals is a widely agreed approach.

*Table 2: Summary of MRI System Setup, Calibration, and Phantom recommendations in a multi-center study.*

| **Summary of MRI System Setup, Calibration, and Phantom recommendations** |
|---|
| $B_0$ has to be the same |
| MRI system capabilities do not have to be the same |
| RF coils do not have to be identical but have to be of the same geometric design |
| Site qualification and QA has to be done, and using static phantoms is the preferred method |
| Calibration methods do not have to be the same<br><br>Viable and preferred calibration methods include<br>● RF power: automatic and manual optimization<br>● Center frequency: water $^{1}$H frequency<br>● Receiver gains: previous *in vivo* measurements<br>● Shimming: automatic shimming on $^{1}$H signals |

# Acquisition and Reconstruction

This section covers topics on acquisition and reconstruction methods for HP [1-$^{13}$C]pyruvate MRI studies. Specialized acquisition and reconstruction methods have been developed for HP [1-$^{13}$C]pyruvate that make careful use of irrecoverable hyperpolarized magnetization, capture the metabolic products of [1-$^{13}$C]lactate, [1-$^{13}$C]alanine, and $^{13}$C-bicarbonate, and can perform rapid imaging to capture the *in vivo* perfusion and metabolism kinetics prior to loss of hyperpolarization due to relaxation. Key consensus recommendations are summarized in Table 3.

## Acquisition

### General acquisition

There was consensus that it is preferred to use the same pulse sequence type when aggregating data across sites (25/26, 96%), where the 3 main pulse sequence types used in HP $^{13}C$ are Magnetic Resonance Spectroscopy Imaging (MRSI), chemical shift encoding (e.g. Iterative Decomposition of water and fat with Echo Asymmetry and Least-squares estimation IDEAL, Chemical Shift Imaging CSI), and metabolite-specific imaging (2). If the same pulse sequence type is used, there was consensus preference that all sites should harmonize sequence parameters (e.g. flip angle scheme, spatial resolution, timing parameters, and undersampling method).

In discussion, there was recognition that different sequences may need to be used within a multi-center study due to variability in available methods and accessible parameter ranges. It was noted that the recommendations were based on the premise that different sequences and parameters can create variability and artifacts, and that comparison of different sequences can be challenging given that the sensitivity and specificity of different sequences is unknown for a specific disease type.

Whether or not the same sequence was used, the group reached consensus that for $^{13}C$ studies it is preferred to use a sequence that provides spatial localization, spectral encoding, and dynamic (time resolved) information, and which captures the bolus of pyruvate.

There was consensus on several standardizing best practice statements, where the group agreed that all studies have to: perform any other MRI contrast injections after the HP $^{13}C$ acquisition (24/25, 96%), since agents such as those including gadolinium will change the HP agent relaxation properties (8); acquire $^{1}H$ images with coverage equal to or exceeding the $^{13}C$ HP FOV (26/26, 100%); and acquire a $B_0$ field map to potentially correct for artifacts and assess $^{13}C$ data quality (25/26, 96%).

### Organ-specific best practice statements

For organ-specific statements, the group members were asked to vote on best practice statements for acquisition methods. It is worth noting that the total number of participating members was smaller (14-20) in this subsection because most members' experiences were limited to one or two organ areas.

Across the **prostate, brain, and abdomen**, there was consensus for best practices to use "metabolite-specific RF excitation" methods (13/15, 87% for prostate, 17/19, 89% for brain, 17/20, 95% for abdomen), for example metabolite-specific flip angles or metabolite-specific excitation. There was also consensus for methods that 'resolve a spectrum' (13/15, 87% for prostate, 14/19, 74% for brain, 15/20, 75% for abdomen), although this could be interpreted to be MRSI (9,10), chemical shift encoding (e.g. IDEAL CSI (11)), and metabolite-specific imaging (12) as they all resolve the spectral dimension somehow. There was no consensus for the use of multi-shot readouts (due to concerns relating to potential loss of pyruvate magnetization; and lack of comparative studies).

Best practice statements for **abdomen** $^{13}C$ studies also include single-shot readouts (13/19, 68%); and respiratory gating methods (18/20, 90%) such as triggering or breath-holding.

Best practice statements for **cardiac** $^{13}C$ studies includes respiratory (13/15, 87%) and cardiac gating (14/15, 93%). There was strong support for metabolite-specific RF excitation (12/14, 86%), to resolve a spectrum (10/14, 71%), and single-shot readouts (12/14, 86%), but these did not meet the predetermined criteria of having at least 15 votes needed to make a best practice statement.

## Reconstruction

The consensus group recommended that within a multi-center study a standardized reconstruction pipeline also be used (25/25, 100%), although this requires all sites use the same acquisition method. It is likely that a standardized pipeline will reduce variation in the quantitative metrics derived from the HP $^{13}C$ signal. However, it was acknowledged in discussion that we have not characterized how variations in the reconstruction pipeline affect resulting metrics

One best practice statement for the reconstruction pipeline reached consensus: using spatial distortion correction (e.g. EPI displacement and ghosting artifact corrections, spiral off-resonance correction) when possible (22/23, 96%).

Lack of comparative evidence made it difficult to reach consensus on a more detailed set of best practice reconstruction statements. No consensus was reached for whether to recommend sum-of-squares or data-driven approaches (13) to combine multi-channel data, and neither was there consensus on using denoising techniques or using zero-filling/spatial filtering/spectral filtering prior to analysis.

Consensus was reached by the group to recommend a minimum set of parameters related to the acquisition and reconstruction that have to be recorded and reported by HP [1-$^{13}$C]pyruvate multi-center human studies, which are shown in Table 5.

*Table 3: Summary of Acquisition and Reconstruction recommendations.*

| **Summary of Acquisition and Reconstruction recommendations** |
|---|
| Multi-center studies are encouraged (where possible) to use the same pulse sequence and to harmonize sequence parameters |
| Spatial, spectral and dynamic information is preferred to be acquired |
| Acquisition strategy recommendations were variable based on body area |
| A standardized reconstruction pipeline is encouraged where the same pulse sequence is used across sites. |

# Data Analysis and Quantification

This section covers topics on data analysis and quantification for HP [1-$^{13}$C]pyruvate MRI studies. HP [1-$^{13}$C]pyruvate *in vivo* experiences perfusion, metabolism, and loss of hyperpolarization due to relaxation, leading to the development of specialized analysis, quantification, and visualization methods. The repeatability and reproducibility of HP [1-$^{13}$C]pyruvate MRI in human subjects is also relatively unexplored so far but could have an important impact on interpreting results in multi-center studies. Key consensus recommendations are summarized in Table 4.

## Data storage/availability

The group discussed standardization of data storage and types within and across multi-center studies. They discussed different types of data a study can acquire and store, including raw data (k-space); minimally processed data (i.e. reconstructed images/spectra); and processed data (parameter maps). There was consensus that all data types were viable and preferred.

The group reached consensus that it is preferred for sites to store all $^{13}$C raw data (e.g. data still in k-space) (26/26, 100%). It was recognized that this will add to the burden of running a multi-center study. However, the group felt that keeping raw data at this early stage of technology development would have significant benefit for development and improvement of acquisition and reconstruction methods, and to ensure consistency and comparability of data.

Consensus was reached by the group to recommend a minimum set of parameters that have to be recorded and reported by HP [1-$^{13}$C]pyruvate multi-center human studies, which are shown in Table 5.

## Data quality/pre-processing

Ensuring high data quality is paramount in clinical studies, as it underpins the validity and reliability of research findings. There was consensus that a data quality assessment has to be made on the results acquired in multi-center HP [1-$^{13}$C]pyruvate studies. SNR measurements, assessment of artifact levels, assessment of dynamic signal curve, and external reference-based assessments were all deemed viable options for data quality assessment, with SNR measurements as the preferred method (26/26, 100%). The choice and type of data quality assessment were recognized to be factors influenced by the specific study design.

Calculation of SNR was discussed further and consensus for viable methods were: temporal peak pyruvate SNR; temporal area-under-curve (AUC) pyruvate SNR; temporal peak total carbon SNR; and temporal AUC total carbon SNR. Of these there was strongest support for the use of temporal AUC total carbon SNR (22/24, 92%). Once again, the group recognized that the choice of SNR calculation methodology may also be determined by specific study design factors.

## Metrics

There was consensus that quantitative evaluation of HP $^{13}$C signals was preferable in multi-center studies. However, quantitative evaluation was not deemed mandatory as it was recognized that studies may be conducted where a qualitative evaluation of data would be useful.

The consensus viable metrics were normalized metabolite signals, metabolite AUCs, normalized metabolite AUCs, metabolite AUC ratios, and kinetic rates, where for HP $^{13}$C the AUC refers to a temporal sum of metabolite signals over time and the kinetic rates are typically modeled based on two-site exchange with a first-order reaction. Metabolite signals alone as a metric did not achieve a consensus. Of these, there was strong preference for metabolite AUC ratios (14) (24/25, 96%) and normalized metabolite AUCs (23/25, 92%). The normalization method for these metrics and the specific kinetic model to use were not specified in the questionnaire, but approaches include use of a reference tissue or a specified parameter such as the maximum pyruvate or average lactate, and these are outlined in our prior position paper (2).

The group recognized that perfusion is likely to be an important factor affecting the signals in a human HP [1-$^{13}$C]pyruvate experiment. There was discussion around whether perfusion measurements, either from the $^{13}$C acquisition or from other methods, should accompany any HP [1-$^{13}$C]pyruvate multi-center study. The group recommended that perfusion assessment be made where possible but recognized in discussion that this might not be achievable for all studies.

## Repeatability and reproducibility

Within this consensus exercise, repeatability refers to an individual site performing repeated hyperpolarized MR measurements using the same equipment while reproducibility refers to the same measurement being made at different sites (allowing for the expected variance in conditions across sites). There was consensus that an assessment of repeatability and reproducibility has to be made in multi-center studies (24/26, 92% for repeatability, 21/25, 84% for reproducibility).

Current viable options that reached consensus for assessing repeatability and reproducibility were the use of static thermally-polarized phantoms, dynamic phantoms that utilize hyperpolarized solutions, and human subjects. The preferred methods were the use of static thermally-polarized phantoms and human subjects. Of these methods, the group recognized in discussion that static phantoms cannot be used to assess many metrics, including AUC ratios and kinetic rate measurements. The use of human subjects was discussed extensively. It was noted that several repeatability assessments of HP [1-$^{13}$C]pyruvate have been reported in relatively small numbers of participants (15,16). It was also noted that reproducibility studies with humans traveling to different sites are logistically challenging and have been reported only once thus far (17). Dynamic phantoms that utilize hyperpolarized solutions were discussed as an alternative method to assess repeatability and reproducibility of final metrics, but the other methods were preferred.

## Visualization

Visualization of human HP [1-$^{13}$C]pyruvate MRI data has included the use of metabolite signal maps, dynamic signal curves, parameter maps, and presentation of anatomical reference images with varying formats and colormap schemes (2). Consensus was reached that all sites in a multi-center HP [1-$^{13}$C]pyruvate MRI study have to use the same visualization methodology when evaluating data (19/25, 76%). Best practice statements that achieved consensus included: acquiring anatomical reference images; displaying metabolite maps overlaid on anatomy; displaying images without masking to regions-of-interests (ROIs); and allowing for use of transparency in $^{13}$C maps overlaid on anatomical images.  No consensus was reached to recommend viewing metabolite maps next to anatomy or using interpolation as best practices.

*Table 4: Summary of Data Analysis and Quantification recommendations.*

| **Summary of Data Analysis and Quantification recommendations** |
| --- |
| Prefer that sites store all $^{13}$C raw data (e.g. in k-space) in addition to reconstructed images/spectra |
| A data quality assessment has to be made on the results acquired in multi-center studies |
| Normalized metabolite AUCs and metabolite AUC ratios are preferred as metrics for assessment of HP [1-$^{13}$C]pyruvate data |
| Repeatability and reproducibility assessments have to be established for multi-center studies, and can be done with static phantoms and/or human subjects |

*Table 5: Minimum set of parameters that have to be recorded and reported in a multi-center HP [1-13C]pyruvate study that had a consensus recommendation.*

| **Hyperpolarized Agent Parameters to be recorded and reported** |
| --- |
| Pyruvate Concentration |
| EPA Concentration |
| Polarization* (*preferred but not required*) |
| pH |
| Temperature |
| Elapsed time from dissolution to start of injection |
| Pyruvate dose injection rate and duration |
| Volume of pyruvate dose injected |
| Saline flush injection rate and duration (if used) |
| **Acquisition Parameters to be recorded and reported** |
| TR, TE, FOV, matrix size, bandwidth, and other ‘standard’ acquisition parameters |
| Timing of acquisition relative to the pyruvate injection |
| Acquisition timing parameters of interval between time points and total number of time points |
| Frequency response specifications and frequency offsets for each metabolite (if a spectrally-selective RF pulse is used) |
| Specified flip angles for each metabolite and each timepoint |
| Echo-spacing, number of echoes, and metabolite frequencies used in the reconstruction (if a multi-echo readout is used) |

# Discussion

Whilst we focused on [1-$^{13}$C]pyruvate preparation for current clinical DNP technology; the majority of the consensus (sections on MR system setup, calibration and phantoms; acquisition and reconstruction; and data analysis and quantification) is equally relevant to competing emerging alternative clinical technologies (e.g. ParaHydrogen-induced polarization (PHIP), Signal Amplification By Reversible Exchange (SABRE) etc). The results from this formal consensus exercise revealed areas with consensus that are largely supported by experimental results as well as areas where there were disagreements and lack of consensus for how to perform HP [1-$^{13}$C]pyruvate MRI multi-center human studies. For HP $^{13}$C-pyruvate preparation, the strongest areas of agreement were that different sites can use different processes, but must use the same release criteria. For MRI system setup, calibration, and phantoms, the group generally recommended that the MRI capabilities, RF coils, and methods do not need to be the same, although it was preferred. The exception was that $B_0$ has to be the same. There was consensus on the use of static phantoms for site qualification and QA, and consensus that there are multiple viable methods for calibration procedures. For acquisition and reconstruction, there was a preference for multi-center studies to use the same pulse sequence and parameters, but this did not reach consensus as a requirement. The group felt that the specific sequence, parameters, and reconstruction should be adapted to the anatomic target and clinical question. For data analysis and quantification, the group recommended that sites in a multi-center study have to perform data quality, repeatability, and reproducibility assessments, where the study would have to define the assessment method.

There was a consensus requirement to store all $^{13}$C raw data. For metrics, SNR was recommended for assessing data quality, but there was no clear preference for the specific output metric(s) to assess metabolism. There was a set of agent and acquisition parameters that we agreed have to be recorded and reported in HP [1-$^{13}$C]pyruvate MRI multi-center human studies.

We believe that this process revealed that HP [1-$^{13}$C]pyruvate MRI as a technology has progressed sufficiently to plan multi-center studies. These studies are critical both as a necessary advancement beyond single site studies but also as a way to achieve larger sample sizes for greater statistical power, enabling assessments as to whether this technology can provide clinically relevant information for patients and clinicians. The consensus building revealed that we largely agree on how HP [1-$^{13}$C]pyruvate should be prepared for multi-center studies, and that there are several viable methods for MRI scanner setup, the acquisition and reconstruction strategy, and analysis of the data.  Given the broad implications of metabolism in disease, but the still limited resources (e.g. polarizers, necessary equipment and expertise), there is still an open question as to which clinical applications will be first to transition into a multi-center study.

The consensus group members participating in this exercise represented 13 different academic institutions.  For each of the topic areas, at least 2 experts in that area led the creation of the consensus questions.  The invited participants were also chosen to include experts from the different topic areas, with particular emphasis to include experts who have developed and overseen the HP $^{13}$C-pyruvate preparation process that is outside the experience of many of the MRI researchers involved in these studies.  The final consensus questionnaire scoring and discussion did require attendance at the face-to-face meeting at UCSF, which prevented the participation of some invitees.

During the process, the number of statements achieving consensus increased from 54.1% in the first round (online) to 62.1% in the second round (face-to-face) (Supplementary Table 3). Eleven statements were removed by the group as being irrelevant, and the language in 18 statements was modified for clarity (Supplementary Table 4).  Some notable areas where consensus was reached only after the second round included the recommendations that sites do not have to follow the same agent preparation style, that reference $^{1}$H images must be acquired without repositioning, and that repeatability and reproducibility assessments have to be established.

## HP $^{13}$C-Pyruvate Preparation

We had numerous discussions about how the regulatory environment and requirements for the HP $^{13}$C-pyruvate preparation processes for human studies would interact with this consensus exercise.  For example, the use of primarily two production styles, Aseptic Preparation and Terminal Sterilization, has in part arisen from variations in local regulatory requirements.  There was the concern that whatever is recommended by this exercise would ultimately be superseded by regulatory requirements. However, it was also noted that HP $^{13}$C researchers are also defining how this technology can be safely and effectively applied in human studies, so there may also be an opportunity to work with regulators to create standards for HP agent preparation that can be broadly used.

The need for the measurement of polarization, both as a release criterion and for reporting data, created significant discussion at the meeting. The discussion primarily centered on the lack of confidence in current approaches for polarization measurements within the HP MRI workflow, whether using the agent in liquid-state after dissolution or measuring in the solid-state prior to dissolution. Reflecting on this result, we realized the intent of the polarization measurement is for efficacy, in other words “will this dose give sufficient signal?”. To potentially resolve this discussion, the current measurement methods likely have room for improvement, for example using both the build-up rate and amplitude in the solid-state to have confidence in the likely efficacy of the resulting dose.

## MRI System Setup, Calibration, and Phantoms

Systems from current major MRI system vendors (GE, Siemens, Philips) all reached consensus as being viable for inclusion in HP human studies, but participants noted some differences in discussions. In particular, GE Healthcare scanners have been used in the majority of studies (2), and so far GE has invested the most into supporting HP studies, including providing pulse sequences tailored for HP applications, which group members noted was a current limitation on other vendors. The use of “vendor-neutral” or 3rd party MR acquisition frameworks such as Pulseq (18), RTHawk (19), or gammaSTAR (20) may offer a solution to support MRI systems from various vendors and standardize pulse sequence methods.

For $^{13}C$ RF coils, the MRI scanners that have been used so far for HP studies did not have $^{13}C$ body coils available, thus requiring use of more-localized $^{13}C$ transmit coils in the patient bore (2). Typically, the transmit $^{13}C$ coils used have larger transmit B1 inhomogeneity compared to the body coil designs used for $^{1}H$ excitation. The uniformity of excitation can have a direct impact on some HP 13C metrics, for example the applied flip angle has a clear influence on the AUC ratio (14) and kinetic rate constants (21). Improving the uniformity transmit $^{13}C$ RF coils would then likely improve the repeatability and reproducibility of these metrics, making such improvements a valuable opportunity for improving multi-center study data.

Phantoms were discussed extensively in the context of site qualification, QA, calibration, repeatability, and reproducibility measurements, highlighting their importance in multi-center studies. Current studies use a variety of phantoms (2) but no standardized phantoms exist. Most phantoms used also do not capture the dynamics and/or metabolic conversion that occurs following HP [1-$^{13}$C]pyruvate injection. This suggests future work is needed on phantom development, including moving towards standardized phantoms, to improve multi-center studies.

## Acquisition and Reconstruction

HP [1-$^{13}$C]pyruvate has been used to image many anatomical regions and different diseases, and the group had extensive discussions about how the recommendations would vary based on the clinical question. It was generally felt that there would be differences in the best options for RF coils, pulse sequence parameters, and analysis methods depending on the anatomical target. It was also generally agreed that the hypothesis of the clinical study would affect the choice of best analysis method, and possibly the acquisition as well. There were some brief discussions of anatomic region differences in the acquisition topic area, but this was limited and also had fewer votes given not all participants had experience in the different regions. Future discussions are needed when a specific anatomical region and/or hypotheses are being proposed for a multi-center study.

## Data Analysis and Quantification

The metrics used for HP [1-$^{13}$C]pyruvate studies are diverse, including metabolite amplitudes, AUCs, kinetic rates, and various types of normalization (2,22). This is very much an active area of research and development. However, for multi-center studies, the group preferred to have a quantitative metric for evaluation, which would require a unified processing pipeline to create this metric. As more HP [1-$^{13}$C]pyruvate human studies are completed, we encourage the community to consider investigation of metrics for metabolism quantification that could be standardized. Ultimately, a standardized metric or metrics along with standardized quantification software would benefit the compilation of data across sites both within and outside of multi-center studies.

While we agreed that repeatability and reproducibility assessments must be performed, we did not clearly conclude how to perform these assessments.  In general, participants wanted to have more data on repeatability and reproducibility, but there are not many examples in human studies.  However, there are more repeatability and reproducibility results from preclinicial imaging studies, and perhaps these could be leveraged to design assessments for multi-center human studies.

# Conclusion

In this consensus building exercise, we examined the entire process required for HP [1-$^{13}$C]pyruvate MRI human studies, including agent preparation, MRI system and hardware, calibration methods, acquisition methods, reconstruction methods, analysis metrics, data visualization, and data storage.  Across these areas, we reached consensus for approximately two-thirds of the statements examined, and that consensus was enhanced through two rounds of voting and a face-to-face meeting.  We hope that these results can be used to support the planning of multi-center HP [1-$^{13}$C]pyruvate MRI human studies, as well as reveal where more work is needed to advance this novel metabolic imaging technology.

## Acknowledgements

“HP $^{13}$C MRI Consensus Group” members:

Aarhus University: Lotte Bonde Bertelsen, Nikolaj Bøgh, Esben Søvsø Szocska Hansen, Christoffer Laustsen, Jack Miller, Michael Vaeggemose

Akroswiss AG: Jonas Steinhauser

Chang Gung Memorial Hospital: Ching-Yi Hsieh, Gigin Lin

ETH Zurich: Max Fuetterer, Sebastian Kozerke

GE Healthcare: Albert Chen, Arnaud  Comment, Adam Gaunt, Justin Lau, Avi Leftin, Rolf Schulte, Jean-Luc Vanderheyden (JLVMI Consulting)

IBEC Barcelona: Irene Marco Rius

Kiel University: Jan-Bernd Hoevener

Kings College London: Thomas C Booth

Laboratory for Functional & Metabolic Imaging (LIFMET): Mor Mishkovsky

MD Anderson Cancer Center: Christopher Michael Walker, Jim Bankson

Memorial Sloan Kettering Cancer Center: Kayvan Keshari

MIAAI (Medical Image Analysis and AI): Ramona Woitek

National Institutes for Quantum Science and Technology, Japan: Yuhei Takado

National Institutes Health/National Cancer Institute: Crystal Vasquez

Nottingham University: Sebastien Serres

NVision Imaging Technologies GmbH: Christoph A. Müller

Oxford Instruments: Joel Floyd

Rapid Biomedical: Titus Lanz

Technical University of Denmark (DTU): Jan Henrik Ardenkjær-Larsen, Andrea Capozzi, Mathilde Hauge Lerche

Technical University of Munich: Martin Grashei, Franz Schilling, Geoffrey J Topping, Frits van Heijster

Funding acknowledgements:

The Hyperpolarized MRI Technology Resource Center (NIH P41EB013598) supported this consensus exercise. DV and DX receive research funding from NIH (R01CA262630). SP receives research support from the National Institute of Health and Care Research (NIHR) University College London Hospital Biomedical Research Centre and an NIHR Research Professorship (NIHR 303144). KRK is supported by the NIH/NCI Cancer Center Support Grant P30CA008748 as well as the Center for Molecular Imaging and Bioengineering (CMIB) at MSKCC. VZ receives research support from The Cancer Prevention Research Institute of Texas (RP180404).

Disclosure of competing interests:

DV, PL, and JG receive research grant funding from GE Healthcare. DV, RB, JS and SP are advisors for NVision Imaging Technologies. JAB is on the scientific advisory board for NVision, and he receives research funding from Siemens. JHA is the owner of Polarize, a company that sells instrumentation for hyperpolarization. FG receives research support from GE Healthcare and grant funding from NVision and AstraZeneca. KRK is a founder of Atish Technologies and a member of the scientific advisory boards of Nvision Imaging Technologies, Imaginostics and Mi2. KRK hold patents related to imaging and modulation of cellular metabolism. SP is an advisor for QUIBIM Ltd, and Cheif Medical Officer for Gold Standard Phantoms. CV receives research support from Siemens.

# Supplementary material for “Consensus Meeting Recommendations for Hyperpolarized [1-$^{13}$C]pyruvate MRI Multi-center Human Studies”

The supplementary material includes additional details from the consensus process such as all statements posed to the members, the detailed results of scoring, and information about the participants.

Tables:

1. Information about all group members who participated.
2. Listing of all statements along with a summary of round 2 scoring indicating the most common outcome (Disagreement/Uncertain/Agreement) and whether consensus was reached (Consensus/No consensus).
3. Summary of all statements scoring with outcomes and consensus for round 1 and round 2.
4. Statements altered between round 1 and round 2.

*Supplementary Table 1: Group Members participating in the consensus voting (name, speciality, country/site, years of C13 expertise). Panellist with * contributed to round 1 scoring but were not able to participate in round 2 face-to-face meeting.*

| Name | Specialty | Country/site | Years of expertise |
|---|---|---|---|
| Esben Sovso Szocska | Engineering / Physics | Aarhus University | 10 |
| Chris Laustsen | Engineering / Physics, Pharmacy Production & Manufacturing | Aarhus University | 15 |
| Lotte Bonde Bertelsen | Engineering / Physics, Pharmacy Production & Manufacturing | Aarhus University | 10 |
| Ferdia Gallagher | Clinician | Cambridge University | 18 |
| Ashley Grimmer | Pharmacy Production & Manufacturing | Cambridge University | 6.5 |
| Mary McLean | Engineering / Physics | Cambridge University | 10 |
| Ching-Yi Hsieh | Engineering / Physics | Chang Gung University and Chang Gung Memorial Hospital at Linkou, Taiwan | 6 |
| Sebastian Kozerke * | Engineering / Physics | ETH Zurich | 15 |
| Albert Chen * | Engineering / Physics | GE Healthcare | 20 |
| Adam Gaunt | Engineering / Physics, Other | GE Healthcare | 10 |
| Arnaud Comment | Engineering / Physics, Pharmacy Production & Manufacturing | GE Healthcare | 15 |
| Jim Bankson | Engineering / Physics | MD Anderson Cancer Center | 14 |
| Kayvan Keshari | Engineering / Physics | Memorial Sloan Kettering - Cancer Center | 15 |
| James Grist | Engineering / Physics | Nottingham University & Oxford University | 10 |
| Damian Tyler | Engineering / Physics | Oxford University | 19 |
| Fulvio Zaccagna | Radiologist / Radiographer, Clinician | Oxford University & Cambridge University | 9 |
| Titus Lanz * | Engineering / Physics | RAPID Biomedical GmbH | 20 |
| Jan Henrik Ardenkjær-Larsen | Engineering / Physics | Technical University of Denmark | 25 |
| Mathilde Lerche | Engineering / Physics, Other | Technical University of Denmark | 22 |
| Peder Larson | Engineering / Physics | UCSF | 17 |
| Duan Xu | Engineering / Physics | UCSF | 13 |
| Bob Bok | Clinician | UCSF | 17 |
| Jim Slater | Pharmacy Production & Manufacturing | UCSF | 8 |
| Duy Dang | Pharmacy Production & Manufacturing | UCSF | 2 |
| Adam Autry | Engineering / Physics | UCSF | 8 |
| Rafat Chowdhury | Engineering / Physics | University College London | 7 |
| Richard Hesketh | Radiologist / Radiographer, Clinician | University College London | 6 |
| Shonit Punwani | Engineering / Physics, Radiologist / Radiographer | University College London | 9 |
| Dirk Mayer | Engineering / Physics | University of Maryland | 17 |
| Chuck Cunningham | Engineering / Physics | University of Toronto | 20 |
| Jae Mo Park | Engineering / Physics | UT Southwestern | 14 |
| Vlad Zaha | Radiologist / Radiographer, Clinician | UT Southwestern | 6 |

**Supplementary Table 2**: Summary of round 2 scoring for all statements included. The outcome category (1-3 disagree, 4-6 uncertain or 7-9 agree) was defined by the median score across all non-zero responders. Consensus indicates whether a 66% or more of the non-zero responses were in the outcome category when at least 14 votes were made.

| No. | Statement | Outcome (round 2) | Consensus (round 2) | Fraction | Percentage |
|---|---|---|---|---|---|
| **1A-1: For multi-center studies:** | | | | | |
| **1** | a. all sites have to follow the same 13C pyruvate preparation guidelines (e.g. Sterile Compounding or Terminal Sterilization). | Disagreement | Consensus | 17/25 | 68% |
| **2** | b. it is preferred that all sites follow the same 13C pyruvate preparation guidelines. | Agreement | No consensus | 15/25 | 60% |
| **1A-2: For multi-center studies:** | | | | | |
| **3** | a. all sites have to follow the same SOPs for 13C pyruvate preparation process. | Disagreement | Consensus | 20/26 | 77% |
| **4** | b. it is preferred that all sites follow the same SOPs for 13C pyruvate preparation process. | Agreement | No consensus | 15/26 | 58% |
| **1A-3: For multi-center studies:** | | | | | |
| **5** | a. all sites have to have the same environment and facilities (e.g. clean room, isolator, clean bench) for 13C pyruvate preparation process. | Disagreement | Consensus | 19/26 | 73% |
| **6** | b. it is preferred that all sites have the same environment and facilities for 13C pyruvate preparation process. | Agreement | No consensus | 13/26 | 50% |
| **1B-1: For multi-center studies:** | | | | | |
| **7** | a. all sites have to follow the same release criteria for release of hyperpolarized 13C pyruvate. | Agreement | Consensus | 21/26 | 81% |
| **8** | b. it is preferred that all sites follow the same release criteria for release of hyperpolarized 13C pyruvate. | Agreement | Consensus | 25/26 | 96% |
| **1B-2: The following have to be part of the dose release criteria:** | | | | | |
| **9** | a. PA concentration | Agreement | Consensus | 25/26 | 96% |
| **10** | b. EPA concentration | Agreement | Consensus | 26/26 | 100% |
| **11** | c. pH | Agreement | Consensus | 26/26 | 100% |
| **12** | d. Temperature | Agreement | Consensus | 23/26 | 88% |
| **13** | e. Volume | Agreement | Consensus | 21/26 | 81% |
| **14** | f. Polarization | Agreement | No consensus | 14/25 | 56% |
| **1B-3: For multi-center studies:** | | | | | |
| **15** | a. all sites have to use the same QC system measurement for dose release. | Disagreement | Consensus | 17/24 | 71% |
| **16** | b. it is preferred that all sites use the same QC system measurement for dose release. | Uncertain | No consensus | 7/26 | 27% |
| **2A-1: For multi-center studies:** | | | | | |
| **17** | a. all sites have to use a scanner from the same manufacturer. (If you are a direct employee of any of the manufacturers listed below, please select "0") | Disagreement | Consensus | 19/22 | 86% |
| **18** | b. it is preferred that all sites use a scanner from the same manufacturer. | Disagreement | Consensus | 17/23 | 74% |
| **2A-2: Viable options for participation in multi-center studies are: (If you are a direct employee of any of the manufacturers listed below, please select "0")** | | | | | |
| **19** | a. GE | Agreement | Consensus | 22/22 | 100% |
| **20** | b. Siemens | Agreement | Consensus | 21/22 | 95% |
| **21** | c. Phillips | Agreement | Consensus | 17/19 | 89% |
| **2A-3: Preferred manufacturer is: (If you are a direct employee of any of the manufacturers listed below, please select "0")** | | | | | |
| **22** | a. GE | Agreement | No consensus | 13/20 | 65% |
| **23** | b. Siemens | Uncertain | No consensus | 10/20 | 50% |
| **24** | c. Phillips | Uncertain | No consensus | 10/19 | 53% |
| **2A-4: For multi-center studies:** | | | | | |
| **25** | a. the magnet field strength B0 has to be the same for all sites. | Agreement | Consensus | 19/24 | 79% |
| **26** | b. it is preferred that all sites use a scanner with the same B0. | Agreement | Consensus | 23/24 | 96% |
| **2A-5: Viable options for participation in multi-center study are:** | | *Question was removed* | | | |
| **27** | a. 0.5T | | | | |
| **28** | b. 1.5T | | | | |
| **29** | c. 3T | | | | |
| **30** | d. 7T | | | | |
| **2A-6: Preferred B0 is:** | | | | | |
| **31** | a. 0.5T | Disagreement | Consensus | 19/23 | 83% |
| **32** | b. 1.5T | Uncertain | Consensus | 16/24 | 67% |
| **33** | c. 3T | Agreement | Consensus | 23/24 | 96% |
| **34** | d. 7T | Uncertain | No consensus | 11/24 | 46% |
| **2A-7: For multi-center studies:** | | | | | |
| **35** | a. all sites have to have the same maximum gradient strength. | Agreement | No consensus | 13/23 | 57% |
| **36** | b. it is preferred that all sites have the same maximum gradient strength. | Agreement | Consensus | 22/23 | 96% |
| **2A-8: For multi-center studies:** | | | | | |
| **37** | a. all sites have to have the same maximum gradient slew rate. | Agreement | No consensus | 14/23 | 61% |
| **38** | b. it is preferred that all sites have the same maximum gradient slew rate. | Agreement | Consensus | 22/23 | 96% |
| **2A-9: For multi-center studies:** | | | | | |
| **39** | a. all sites have to have the same maximum available RF amplifier power. | Agreement | No consensus | 12/23 | 52% |
| **40** | b. it is preferred that all sites have the same maximum available amplifier power. | Agreement | Consensus | 22/23 | 96% |
| **2A-10: For multi-center studies:** | | | | | |
| **41** | a. all sites have to have the same manufacturer and model for 13C transmit and receive coils. | Disagreement | No consensus | 13/24 | 54% |
| **42** | b. it is preferred that all sites have the same manufacturer and model for 13C transmit and receive coils. | Agreement | Consensus | 20/24 | 83% |
| **2A-11: For multi-center studies:** | | | | | |
| **43** | a. all sites have to have the same fundamental geometry design for 13C transmit and receive coils. | Uncertain | No consensus | 6/23 | 26% |
| **44** | b. it is preferred that all sites have the same fundamental geometry design for 13C transmit and receive coils. | Agreement | Consensus | 22/23 | 96% |
| **2A-12: For multi-center studies:** | | | | | |
| **45** | a. all sites have to have a reference phantom in the field of view of 13C coil. | Uncertain | No consensus | 8/24 | 33% |
| **46** | b. it is preferred that all sites have to have a reference phantom in the field of view of 13C coil. | Agreement | No consensus | 15/24 | 63% |
| **2A-13: For multi-center studies:** | | | | | |
| **47** | a. all sites have to be able to acquire 1H and 13C images without repositioning the patient in between acquisition scans. | Agreement | Consensus | 16/24 | 67% |
| **48** | b. it is preferred that all sites are able to acquire 1H and 13C images without repositioning the patient in between acquisition scans. | Agreement | Consensus | 23/24 | 96% |
| **2B-1: For multi-center studies:** | | | | | |
| **49** | a. all sites have to perform an agreed-upon scanner site qualification (e.g. QC experiments). | Agreement | Consensus | 23/24 | 96% |
| **50** | b. it is preferred that all sites perform an agreed-upon scanner site qualification. | Agreement | Consensus | 24/24 | 100% |
| **2B-2: Viable approaches to scanner site qualification include use of:** | | | | | |
| **51** | a. Static thermal phantoms | Agreement | Consensus | 25/25 | 100% |
| **52** | b. Dynamic HP phantoms (e.g. with enzymatic conversion) | Uncertain | No consensus | 5/25 | 20% |
| **53** | c. Travelling humans injected with HP 13C | Uncertain | No consensus | 8/24 | 33% |
| **2B-3: Preferred approach is:** | | | | | |
| **54** | a. Static thermal phantoms | Agreement | Consensus | 25/25 | 100% |
| **55** | b. Dynamic HP phantoms (e.g. with enzymatic conversion) | Uncertain | No consensus | 11/25 | 44% |
| **56** | c. Travelling humans injected with HP 13C | Disagreement | No consensus | 12/24 | 50% |
| **2B-4: For multi-center studies:** | | | | | |
| **57** | a. all sites have to perform on-going hardware quality assurance throughout the study. | Agreement | Consensus | 20/25 | 80% |
| **58** | b. it is preferred that all sites perform on-going hardware quality assurance throughout the study. | Agreement | Consensus | 24/25 | 96% |
| **2C-1: For multi-center studies:** | | | | | |
| **59** | a. all sites have to follow the same protocol for adjustment of RF pulse power. | Disagreement | No consensus | 11/22 | 50% |
| **60** | b. it is preferred that all sites follow the same protocol for adjustment of RF pulse power. | Agreement | Consensus | 17/22 | 77% |
| **2C-2: Viable options for pre-scan setting of RF pulse power are:** | | | | | |
| **61** | a. Automated optimization, e.g. Bloch-Siegert | Agreement | Consensus | 21/22 | 95% |
| **62** | b. Manual optimization, e.g. power sweep | Agreement | Consensus | 22/22 | 100% |
| **63** | c. Predetermined values from phantom calibrations | Agreement | No consensus | 13/22 | 59% |
| **2C-3: Preferred method of adjustment of RF pulse power:** | | | | | |
| **64** | a. Automated optimization, e.g. Bloch-Siegert | Agreement | Consensus | 21/21 | 100% |
| **65** | b. Manual optimization, e.g. power sweep | Agreement | Consensus | 19/21 | 90% |
| **66** | c. Predetermined values from phantom calibrations | Uncertain | No consensus | 7/21 | 33% |
| **2C-4: For multi-center studies:** | | | | | |
| **67** | a. all sites have to follow the same protocol for pre-scan adjustment of center frequency. | Uncertain | No consensus | 4/24 | 17% |
| **68** | b. it is preferred that all sites follow the same protocol for pre-scan adjustment of center frequency. | Agreement | Consensus | 23/24 | 96% |
| **2C-5: Viable options for pre-scan setting of center frequency are:** | | | | | |
| **69** | a. 1H frequency in tissue | Agreement | Consensus | 24/24 | 100% |
| **70** | b. 13C frequency in phantom | Agreement | Consensus | 17/24 | 71% |
| **71** | c. 13C frequency in tissue | Uncertain | No consensus | 11/24 | 46% |
| **72** | d. 23Na frequency in tissue | Uncertain | No consensus | 12/24 | 50% |
| **2C-6: Preferred method of adjustment of center frequency is:** | | | | | |
| **73** | a. 1H frequency in tissue | Agreement | Consensus | 23/24 | 96% |
| **74** | b. 13C frequency in phantom | Uncertain | No consensus | 9/24 | 38% |
| **75** | c. 13C frequency in tissue | Uncertain | No consensus | 10/24 | 42% |
| **76** | d. 23Na frequency in tissue | Uncertain | No consensus | 11/24 | 46% |
| **2C-7: For multi-center studies:** | | | | | |
| **77** | a. all sites have to follow the same protocol for adjustment of receive gains. | Uncertain | No consensus | 7/21 | 33% |
| **78** | b. it is preferred that all sites follow the same protocol for adjustment of receive gains. | Uncertain | No consensus | 6/21 | 29% |
| **2C-8: Viable options for pre-scan setting of receive gains are:** | | | | | |
| **79** | a. Maximum values possible | Disagreement | No consensus | 11/21 | 52% |
| **80** | b. Based on previous in vivo measurements | Agreement | Consensus | 19/22 | 86% |
| **81** | c. Based on phantom measurements | Agreement | No consensus | 12/22 | 55% |
| **2C-9: Preferred method of adjustment of receive gains is:** | | | | | |
| **82** | a. Maximum values possible | Disagreement | No consensus | 11/21 | 52% |
| **83** | b. Based on previous in vivo measurements | Agreement | Consensus | 19/22 | 86% |
| **84** | c. Based on phantom measurements | Agreement | No consensus | 12/22 | 55% |
| **2C-10: For multi-center studies:** | | | | | |
| **85** | a. all sites have to follow the same protocol for shimming to optimize B0 homogeneity. | Uncertain | No consensus | 7/24 | 29% |
| **86** | b. it is preferred that all sites follow the same protocol for shimming to optimize B0 homogeneity. | Agreement | Consensus | 21/24 | 88% |
| **2C-11: Viable options for shimming are:** | | | | | |
| **87** | a. Automated shimming over imaging volume | Agreement | Consensus | 22/23 | 96% |
| **88** | b. Automated shimming over a ROI (e.g. PRESS Box) | Agreement | Consensus | 21/23 | 91% |
| **89** | c. Manual shimming over imaging volume | Uncertain | No consensus | 6/23 | 26% |
| **90** | d. Manual shimming over a ROI | Agreement | No consensus | 12/23 | 52% |
| **2C-12: Preferred methods of adjustment of shims are:** | | | | | |
| **91** | a. Automated shimming over imaging volume | Agreement | Consensus | 22/23 | 96% |
| **92** | b. Automated shimming over a ROI (e.g. PRESS Box) | Agreement | Consensus | 22/23 | 96% |
| **93** | c. Manual shimming over imaging volume | Uncertain | No consensus | 11/23 | 48% |
| **94** | d. Manual shimming over a ROI | Uncertain | No consensus | 10/23 | 43% |
| **2C-13: For multi-center studies:** | | | | | |
| **95** | a. all sites have to perform higher-order shimming. | Disagreement | No consensus | 15/24 | 63% |
| **96** | b. it is preferred that all sites perform higher-order shimming. | Uncertain | No consensus | 6/24 | 25% |
| **3A-1: For multi-center studies:** | | | | | |
| **97** | a. all sites have to use the same pulse sequence type (e.g. MRS/I, metabolite-specific imaging, or chemical shift encoding) when aggregating data across sites. | Uncertain | No consensus | 10/26 | 38% |
| **98** | b. it is preferred that all sites use the same pulse sequence type when aggregating data across sites. | Agreement | Consensus | 25/26 | 96% |
| **3A-2: For multi-center studies:** | | | | | |
| **99** | a. all sites have to use sequences that provide spatial localization. | Agreement | No consensus | 14/26 | 54% |
| **100** | b. it is preferred that all sites use sequences that provide spatial localization. | Agreement | Consensus | 22/26 | 85% |
| **3A-3: For multi-center studies:** | | | | | |
| **101** | a. all sites have to use sequences that provide spectral information (e.g. generate metabolite maps, provide a spectrum). | Uncertain | No consensus | 9/26 | 35% |
| **102** | b. it is preferred that all sites use sequences that provide spectral information (e.g. generate metabolite maps, provide a spectrum). | Agreement | Consensus | 26/26 | 100% |
| **3A-4: For multi-center studies:** | | | | | |
| **103** | a. all sites have to use sequences that provide dynamic (time resolved) data. | Uncertain | No consensus | 8/26 | 31% |
| **104** | b. it is preferred that all sites use sequences that provide dynamic (time resolved) data. | Agreement | Consensus | 24/26 | 92% |
| **3A-5: For multi-center studies:** | | | | | |
| **105** | a. all sites have to capture the bolus of pyruvate in the data acquisition. | Uncertain | No consensus | 11/26 | 42% |
| **106** | b. it is preferred that all sites capture the bolus of pyruvate in the data acquisition. | Agreement | Consensus | 23/26 | 88% |
| **3A-6: For a given pulse sequence:** | | | | | |
| **107** | a. all sites have to use the same flip angle scheme. | Uncertain | No consensus | 4/26 | 15% |
| **108** | b. it is preferred that all sites use the same flip angle scheme. | Agreement | Consensus | 25/26 | 96% |
| **3A-7: For multi-center studies:** | | | | | |
| **109** | a. all sites have to use the same spatial resolution when aggregating data across sites. | Uncertain | No consensus | 9/26 | 35% |
| **110** | b. it is preferred that all sites use the same spatial resolution when aggregating data across sites. | Agreement | Consensus | 25/26 | 96% |
| **3A-8: For multi-center studies:** | | | | | |
| **111** | a. all sites have to have the same timing parameters (start time, temporal resolution, and number of timeframes) when aggregating data across different sites. | Uncertain | No consensus | 7/26 | 27% |
| **112** | b. it is preferred that all sites have the same timing parameters when aggregating data across different sites. | Agreement | Consensus | 25/26 | 96% |
| **3A-9: For multi-center studies:** | | | | | |
| **113** | a. all sites have to use the same undersampling method (e.g. partial Fourier acceleration, parallel imaging acceleration, model-based acceleration methods such as compressed sensing, low-rank reconstructions, and deep learning). | Uncertain | No consensus | 4/23 | 17% |
| **114** | b. it is preferred that all sites have the same undersampling method (e.g. partial Fourier acceleration, parallel imaging acceleration, model-based acceleration methods such as compressed sensing, low-rank reconstructions, and deep learning). | Agreement | Consensus | 22/23 | 96% |
| **3A-10: Best practice recommendation** | | | | | |
| **115** | a. Any contrast injection needed for 1H MRI have to be performed after the HP 13C study. | Agreement | Consensus | 24/25 | 96% |
| **116** | b. 1H images with coverage equal to or exceeding the 13C FOV have to be acquired, for anatomic reference. | Agreement | Consensus | 26/26 | 100% |
| **117** | c. Acquire a B0 field map to identify and potentially correct for artifacts caused by B0 inhomogeneity | Agreement | Consensus | 25/26 | 96% |
| **3B-1: Best practice recommendation for PROSTATE acquisition includes:** | | | | | |
| **118** | a. Multi-shot readouts | Uncertain | No consensus | 7/15 | 47% |
| **119** | b. Single-shot readouts | Agreement | No consensus | 8/15 | 53% |
| **120** | c. Metabolite-specific RF excitation | Agreement | Consensus | 13/15 | 87% |
| **121** | d. Resolve a spectrum | Agreement | Consensus | 13/15 | 87% |
| **122** | e. Respiratory gating | Disagreement | Consensus | 13/15 | 87% |
| **123** | f. Cardiac gating | Disagreement | Consensus | 14/15 | 93% |
| **3B-2: Best practice recommendation for BRAIN acquisition includes:** | | | | | |
| **124** | a. Multi-shot readouts | Uncertain | No consensus | 7/19 | 37% |
| **125** | b. Single-shot readouts | Agreement | Consensus | 14/19 | 74% |
| **126** | c. Metabolite-specific RF excitation | Agreement | Consensus | 17/19 | 89% |
| **127** | d. Resolve a spectrum | Agreement | Consensus | 14/19 | 74% |
| **128** | e. Respiratory gating | Disagreement | Consensus | 18/20 | 90% |
| **129** | f. Cardiac gating | Disagreement | Consensus | 18/20 | 90% |
| **3B-3: Best practice recommendation for CARDIAC acquisition includes:** | | | | | |
| **130** | a. Multi-shot readouts | Disagreement | No consensus | 8/14 | 57% |
| **131** | b. Single-shot readouts | Agreement | Consensus | 12/14 | 86% |
| **132** | c. Metabolite-specific RF excitation | Agreement | Consensus | 12/14 | 86% |
| **133** | d. Resolve a spectrum | Agreement | Consensus | 10/14 | 71% |
| **134** | e. Respiratory gating | Agreement | Consensus | 13/15 | 87% |
| **135** | f. Cardiac gating | Agreement | Consensus | 14/15 | 93% |
| **3B-4: Best practice recommendation for ABDOMEN acquisition includes:** | | | | | |
| **136** | a. Multi-shot readouts | Uncertain | No consensus | 10/19 | 53% |
| **137** | b. Single-shot readouts | Agreement | Consensus | 13/19 | 68% |
| **138** | c. Metabolite-specific RF excitation | Agreement | Consensus | 17/20 | 85% |
| **139** | d. Resolve a spectrum | Agreement | Consensus | 15/20 | 75% |
| **140** | e. Respiratory gating | Agreement | Consensus | 18/20 | 90% |
| **141** | f. Cardiac gating | Uncertain | No consensus | 8/20 | 40% |
| **3C-1: For multi-center studies:** | | | | | |
| **142** | a. a standardized reconstruction pipeline has to be used when aggregating data. | Agreement | No consensus | 14/25 | 56% |
| **143** | b. it is preferred that a standardized reconstruction pipeline is used when aggregating data. | Agreement | Consensus | 25/25 | 100% |
| **3C-2: Best practice recommendation includes:** | | | | | |
| **144** | a. Using denoising techniques to denoise 13C data. | Uncertain | No consensus | 11/24 | 46% |
| **145** | b. Sum-of-squares combination of multi-channel data. | Uncertain | No consensus | 4/20 | 20% |
| **146** | c. HP data-driven approaches for combination of multi-channel data. | Agreement | No consensus | 13/20 | 65% |
| **147** | d. Zero-filling prior to analysis. | Agreement | No consensus | 10/20 | 50% |
| **148** | e. Spatial filtering prior to analysis. | Uncertain | No consensus | 8/21 | 38% |
| **149** | f. Spectral filtering prior to analysis. | Agreement | No consensus | 12/21 | 57% |
| **150** | g. Spatial distortion correction (e.g. EPI displacement and ghosting artifact corrections, spiral off-resonance correction). | Agreement | Consensus | 22/23 | 96% |
| **3D-1: To facilitate analysis, all sites in a multi-center study have to record the following study parameters:** | | | | | |
| **151** | a. 'Standard' scan parameters, such as TR, TE, FOV, matrix size, and bandwidth need to be reported by each site. | Agreement | Consensus | 25/25 | 100% |
| **152** | b. Timing parameters for the pyruvate injection and flush (start, duration) and acquisition (start : interval : end) need to be reported by each site. | Agreement | Consensus | 24/25 | 96% |
| **153** | c. If a spectrally-selective RF pulse is used, the frequency response specifications and frequency offsets for each metabolite need to be clearly stated by each site. | Agreement | Consensus | 24/25 | 96% |
| **154** | d. If a variable flip angle scheme is used, the flip angles for each metabolite and each timepoint need to be reported by each site. | Agreement | Consensus | 25/25 | 100% |
| **155** | e. If a multi-echo readout is used for chemical shift encoding, the echo-spacing, number of echoes, and metabolite frequencies used in the reconstruction should be reported by each site. | Agreement | Consensus | 25/25 | 100% |
| **4A-1: For multi-center studies:** | | | | | |
| **156** | a. all sites have to store all 13C k-space raw data in addition to reconstructed images/spectra. | Agreement | No consensus | 17/26 | 65% |
| **157** | b. it is preferred that all sites have to store all 13C k-space raw data in addition to reconstructed images/spectra. | Agreement | Consensus | 26/26 | 100% |
| **4A-2: Viable data for storage includes:** | | | | | |
| **158** | a. Raw data (k-space) | Agreement | Consensus | 25/25 | 100% |
| **159** | b. Minimally processed data (reconstructed images/spectra) | Agreement | Consensus | 25/25 | 100% |
| **160** | c. Processed data (parameter maps) | Agreement | Consensus | 24/25 | 96% |
| **4A-3: Preferred data for storage are:** | | | | | |
| **161** | a. Raw data (k-space) | Agreement | Consensus | 25/25 | 100% |
| **162** | b. Minimally processed data (reconstructed images/spectra) | Agreement | Consensus | 24/25 | 96% |
| **163** | c. Processed data (parameter maps) | Agreement | Consensus | 23/25 | 92% |
| **4A-4: For multi-center studies, the following additional data has to be recorded:** | | | | | |
| **164** | a. QC - pyruvate concentration | Agreement | Consensus | 25/28 | 89% |
| **165** | b. QC - EPA concentration | Agreement | Consensus | 23/28 | 82% |
| **166** | c. QC - Polarization | Agreement | No consensus | 17/28 | 61% |
| **167** | d. QC - pH | Agreement | Consensus | 24/28 | 86% |
| **168** | e. QC - temperature | Agreement | Consensus | 21/28 | 75% |
| **169** | f. Volume injected | Agreement | Consensus | 26/28 | 93% |
| **170** | g. Injection rate | Agreement | Consensus | 25/28 | 89% |
| **171** | h. Elapsed time from dissolution to start of injection | Agreement | Consensus | 23/28 | 82% |
| **172** | i. Elapsed time between start of injection and start of data acquisition | Agreement | Consensus | 24/28 | 86% |
| **4A-5: It is preferred that the following additional data are recorded:** | | | | | |
| **173** | a. QC - pyruvate concentration | Agreement | Consensus | 27/28 | 96% |
| **174** | b. QC - EPA concentration | Agreement | Consensus | 26/28 | 93% |
| **175** | c. QC - Polarization | Agreement | Consensus | 26/28 | 93% |
| **176** | d. QC - pH | Agreement | Consensus | 27/28 | 96% |
| **177** | e. QC - temperature | Agreement | Consensus | 25/28 | 89% |
| **178** | f. Volume injected | Agreement | Consensus | 27/28 | 96% |
| **179** | g. Injection rate | Agreement | Consensus | 28/28 | 100% |
| **180** | h. Elapsed time from dissolution to start of injection | Agreement | Consensus | 28/28 | 100% |
| **181** | i. Elapsed time between start of injection and start of data acquisition | Agreement | Consensus | 28/28 | 100% |
| **4A-6: Basic patient metadata to be recorded/stored:** | | *Question was removed* | | | |
| **182** | a. Diagnosis/disease | | | | |
| **183** | b. Disease stage (e.g. TNM status, grade category) | | | | |
| **184** | c. Age | | | | |
| **185** | d. Gender | | | | |
| **186** | e. Ethnicity | | | | |
| **187** | f. Weight and Height | | | | |
| **188** | g. Vital signs on day of study (e.g. blood pressure, heart rate) | | | | |
| **4B-1: For multi-center studies:** | | | | | |
| **189** | a. data quality assessment has to be made. | Agreement | Consensus | 25/26 | 96% |
| **190** | b. it is preferred that data quality assessment to be made. | Agreement | Consensus | 25/26 | 96% |
| **4B-2: Viable method to assess data quality are:** | | | | | |
| **191** | a. SNR Measurements | Agreement | Consensus | 26/26 | 100% |
| **192** | b. Assessment of artifact levels | Agreement | Consensus | 22/26 | 85% |
| **193** | c. Assessment of dynamic signal curves | Agreement | Consensus | 18/25 | 72% |
| **194** | d. External Reference-based assessments | Agreement | Consensus | 21/25 | 84% |
| **4B-3: Preferred method to assess data quality are:** | | | | | |
| **195** | a. SNR Measurements | Agreement | Consensus | 26/26 | 100% |
| **196** | b. Assessment of artifact levels | Agreement | Consensus | 21/26 | 81% |
| **197** | c. Assessment of dynamic signal curves | Agreement | Consensus | 21/25 | 84% |
| **198** | d. External Reference-based assessments | Agreement | No consensus | 15/23 | 65% |
| **4B-4: Viable method for calculating 13C SNR for data quality assessment includes:** | | | | | |
| **199** | a. Temporal peak Pyruvate SNR | Agreement | Consensus | 23/24 | 96% |
| **200** | b. Temporal AUC Pyruvate SNR | Agreement | Consensus | 23/24 | 96% |
| **201** | c. Temporal peak total Carbon SNR | Agreement | Consensus | 23/24 | 96% |
| **202** | d. Temporal AUC total Carbon SNR | Agreement | Consensus | 23/24 | 96% |
| **4B-5: Preferred method for calculating 13C SNR for data quality assessment are:** | | | | | |
| **203** | a. Temporal peak Pyruvate SNR | Agreement | No consensus | 15/24 | 63% |
| **204** | 4b. Temporal AUC Pyruvate SNR | Agreement | Consensus | 19/24 | 79% |
| **205** | c. Temporal peak total Carbon SNR | Agreement | Consensus | 18/24 | 75% |
| **206** | d. Temporal AUC total Carbon SNR | Agreement | Consensus | 22/24 | 92% |
| **4C-1: For multi-center studies:** | | | | | |
| **207** | a. HP MRI data has to be evaluated with quantitative metabolism metrics. | Agreement | No consensus | 13/26 | 50% |
| **208** | b. it is preferred that HP MRI data to be evaluated with quantitative metabolism metrics. | Agreement | Consensus | 25/26 | 96% |
| **4C-2: Viable options for evaluating HP MRI data include calculation of:** | | | | | |
| **209** | a. Metabolite signals | Agreement | No consensus | 16/25 | 64% |
| **210** | b. Normalized Metabolite signals | Agreement | Consensus | 23/25 | 92% |
| **211** | c. Metabolite AUCs | Agreement | Consensus | 21/25 | 84% |
| **212** | d. Normalized Metabolite AUCs | Agreement | Consensus | 23/25 | 92% |
| **213** | e. Metabolite AUC ratios | Agreement | Consensus | 25/25 | 100% |
| **214** | f. Kinetic rates | Agreement | Consensus | 23/25 | 92% |
| **4C-3: The preferred metric for evaluating HP MRI data is:** | | | | | |
| **215** | a. Metabolite signals | Agreement | No consensus | 13/26 | 50% |
| **216** | b. Normalized Metabolite signals | Agreement | Consensus | 18/26 | 69% |
| **217** | c. Metabolite AUCs | Agreement | Consensus | 17/25 | 68% |
| **218** | d. Normalized Metabolite AUCs | Agreement | Consensus | 23/25 | 92% |
| **219** | e. Metabolite AUC ratios | Agreement | Consensus | 24/25 | 96% |
| **220** | f. Kinetic rates | Agreement | Consensus | 19/25 | 76% |
| **4C-4: Multi-center HP MRI studies:** | | | | | |
| **221** | a. have to incorporate a measurement of tissue perfusion to explore as a covariate with other metrics. | Uncertain | No consensus | 7/25 | 28% |
| **222** | b. it is preferred that multi-center HP MRI studies incorporate a measurement of tissue perfusion to explore as a covariate with other metrics. | Agreement | Consensus | 20/25 | 80% |
| **4D-1: For multi-center studies:** | | | | | |
| **223** | a. an assessment of repeatability has to be made. | Agreement | Consensus | 24/26 | 92% |
| **224** | b. it is preferred that an assessment of repeatability to be made. | Agreement | Consensus | 26/26 | 100% |
| **4D-2: Viable options for assessing repeatability include use of:** | | | | | |
| **225** | a. Static thermal phantoms | Agreement | Consensus | 22/25 | 88% |
| **226** | b. Dynamic phantoms (e.g. enzymatic conversion) | Agreement | No consensus | 16/25 | 64% |
| **227** | c. Same Subject injected with repeat HP 13C injections | Agreement | Consensus | 21/25 | 84% |
| **4D-3: Preferred options for assessing repeatability include use of:** | | | | | |
| **228** | a. Static thermal phantoms | Agreement | Consensus | 20/25 | 80% |
| **229** | b. Dynamic phantoms (e.g. enzymatic conversion) | Agreement | No consensus | 14/25 | 56% |
| **230** | c. Same Subject injected with repeat HP 13C injections | Agreement | Consensus | 21/25 | 84% |
| **4D-4: For multi-center studies:** | | | | | |
| **231** | a. an assessment of reproducibility has to be made. | Agreement | Consensus | 21/25 | 84% |
| **232** | b. it is preferred that an assessment of reproducibility to be made. | Agreement | Consensus | 25/25 | 100% |
| **4D-5: Viable options for assessing reproducibility include use of:** | | | | | |
| **233** | a. Static thermal phantoms | Agreement | Consensus | 22/25 | 88% |
| **234** | b. Dynamic phantoms (e.g. enzymatic conversion) | Agreement | Consensus | 18/25 | 72% |
| **235** | c. Same Subject injected with HP 13C injections at more than one site | Agreement | Consensus | 20/25 | 80% |
| **4D-6: Preferred options for assessing reproducibility include use of:** | | | | | |
| **236** | a. Static thermal phantoms | Agreement | Consensus | 18/25 | 72% |
| **237** | b. Dynamic phantoms (e.g. enzymatic conversion) | Agreement | No consensus | 14/25 | 56% |
| **238** | c. Same Subject injected with HP 13C injections at more than one site | Agreement | Consensus | 19/25 | 76% |
| **4E-1: Best practice for visualizing hyperpolarized MR data includes:** | | | | | |
| **239** | a. Anatomical reference image required | Agreement | Consensus | 24/25 | 96% |
| **240** | b. Metabolite maps overlaid on anatomy | Agreement | Consensus | 23/25 | 92% |
| **241** | c. Metabolite maps next to anatomy | Agreement | No consensus | 16/25 | 64% |
| **242** | d. Images without masking to ROIs | Agreement | Consensus | 19/25 | 76% |
| **243** | e. Use of transparency in overlaid maps | Agreement | Consensus | 22/25 | 88% |
| **244** | f. Interpolation | Uncertain | No consensus | 11/25 | 44% |
| **4E-2: For multi-center studies:** | | | | | |
| **245** | a. all sites have to use the same visualization methodology when evaluating data. | Agreement | Consensus | 19/25 | 76% |
| **246** | b. it is preferred that all sites to use the same visualization methodology when evaluating data. | Agreement | Consensus | 25/25 | 100% |

*Supplementary Table 3: Summary of how all statements scored by participants during Round 1 and Round 2.*

| Consensus round | Agreement with consensus, *n* (%) | Disagreement with consensus, *n* (%) | Uncertainty or no consensus, *n* (%) |
|---|---|---|---|
| Round 1 (*n*=246) | 133 (54.1%) | 3 (1.2%) | 110 (44.7%) |
| Round 2 (*n*=235) | 146 (62.1%) | 11 (4.7%) | 74 (33.2%) |

*Supplementary Table 4: Summary of statement additions, deletions, and changes made after round 1 and during round 2.*

| Altered statement post- round 2 | Previous statement | New statement |
|---|---|---|
| | 2A-5: Viable options for participation in multi-center study are: | Removed |
| **27** | a. 0.5T | Removed |
| **28** | b. 1.5T | Removed |
| **29** | c. 3T | Removed |
| **30** | d. 7T | Removed |
| | 4A-4: For multi-center studies, the following additional data has to be **recorded**: | 4A-4: For multi-center studies, the following additional data has to be **recorded and reported**: |
| **164** | a. QC - pyruvate concentration | a. QC - pyruvate concentration |
| **165** | b. QC - EPA concentration | b. QC - EPA concentration |
| **166** | c. QC - Polarization | c. QC - Polarization |
| **167** | d. QC - pH | d. QC - pH |
| **168** | e. QC - temperature | e. QC - temperature |
| **169** | f. Volume injected | f. Volume injected |
| **170** | g. Injection rate | g. Injection rate |
| **171** | h. Elapsed time from dissolution to start of injection | h. Elapsed time from dissolution to start of injection |
| **172** | i. Elapsed time between start of injection and start of data acquisition | i. Elapsed time between start of injection and start of data acquisition |
| | 4A-5: It is preferred that the following additional data are **recorded**: | 4A-5: It is preferred that the following additional data are **recorded and reported**: |
| **173** | a. QC - pyruvate concentration | a. QC - pyruvate concentration |
| **174** | b. QC - EPA concentration | b. QC - EPA concentration |
| **175** | c. QC - Polarization | c. QC - Polarization |
| **176** | d. QC - pH | d. QC - pH |
| **177** | e. QC - temperature | e. QC - temperature |
| **178** | f. Volume injected | f. Volume injected |
| **179** | g. Injection rate | g. Injection rate |
| **180** | h. Elapsed time from dissolution to start of injection | h. Elapsed time from dissolution to start of injection |
| **181** | i. Elapsed time between start of injection and start of data acquisition | i. Elapsed time between start of injection and start of data acquisition |
| | 4A-6: Basic patient metadata to be recorded/stored: | Removed |
| **182** | a. Diagnosis/disease | Removed |
| **183** | b. Disease stage (e.g. TNM status, grade category) | Removed |
| **184** | c. Age | Removed |
| **185** | d. Gender | Removed |
| **186** | e. Ethnicity | Removed |
| **187** | f. Weight and Height | Removed |
| **188** | g. Vital signs on day of study (e.g. blood pressure, heart rate) | Removed |